\newif\ifpagetitre            \pagetitretrue
\newtoks\hautpagetitre        \hautpagetitre={\hfil}
\newtoks\baspagetitre         \baspagetitre={\hfil}
\newtoks\auteurcourant        \auteurcourant={\hfil}
\newtoks\titrecourant         \titrecourant={\hfil}

\newtoks\hautpagegauche       \newtoks\hautpagedroite
\hautpagegauche={\hfil\the\auteurcourant\hfil}
\hautpagedroite={\hfil\the\titrecourant\hfil}

\newtoks\baspagegauche \baspagegauche={\hfil\tenrm\folio\hfil}
\newtoks\baspagedroite \baspagedroite={\hfil\tenrm\folio\hfil}

\headline={\ifpagetitre\the\hautpagetitre
\else\ifodd\pageno\the\hautpagedroite
\else\the\hautpagegauche\fi\fi}

\footline={\ifpagetitre\the\baspagetitre
\global\pagetitrefalse
\else\ifodd\pageno\the\baspagedroite
\else\the\baspagegauche\fi\fi}

\vsize=9.0in\voffset=1cm
\looseness=2


\message{fonts,}

\font\tenrm=cmr10
\font\ninerm=cmr9
\font\eightrm=cmr8
\font\teni=cmmi10
\font\ninei=cmmi9
\font\eighti=cmmi8
\font\ninesy=cmsy9
\font\tensy=cmsy10
\font\eightsy=cmsy8
\font\tenbf=cmbx10
\font\ninebf=cmbx9
\font\tentt=cmtt10
\font\ninett=cmtt9

\font\ninesl=cmsl9
\font\eightsl=cmsl8

\font\nineit=cmti9
\font\eightit=cmti8

\skewchar\ninei='177 \skewchar\eighti='177
\skewchar\ninesy='60 \skewchar\eightsy='60

\def\eightpoint{\def\rm{\fam0\eightrm} 
\normalbaselineskip=9pt
\normallineskiplimit=-1pt
\normallineskip=0pt

\textfont0=\eightrm \scriptfont0=\sevenrm \scriptscriptfont0=\fiverm
\textfont1=\ninei \scriptfont1=\seveni \scriptscriptfont1=\fivei
\textfont2=\ninesy \scriptfont2=\sevensy \scriptscriptfont2=\fivesy
\textfont3=\tenex \scriptfont3=\tenex \scriptscriptfont3=\tenex
\textfont\itfam=\eightit  \def\it{\fam\itfam\eightit} 
\textfont\slfam=\eightsl \def\sl{\fam\slfam\eightsl} 

\setbox\strutbox=\hbox{\vrule height6pt depth2pt width0pt}%
\normalbaselines \rm}

\def\ninepoint{\def\rm{\fam0\ninerm} 
\textfont0=\ninerm \scriptfont0=\sevenrm \scriptscriptfont0=\fiverm
\textfont1=\ninei \scriptfont1=\seveni \scriptscriptfont1=\fivei
\textfont2=\ninesy \scriptfont2=\sevensy \scriptscriptfont2=\fivesy
\textfont3=\tenex \scriptfont3=\tenex \scriptscriptfont3=\tenex
\textfont\itfam=\nineit  \def\it{\fam\itfam\nineit} 
\textfont\slfam=\ninesl \def\sl{\fam\slfam\ninesl} 
\textfont\bffam=\ninebf \scriptfont\bffam=\sevenbf
\scriptscriptfont\bffam=\fivebf \def\bf{\fam\bffam\ninebf} 
\textfont\ttfam=\ninett \def\tt{\fam\ttfam\ninett} 

\normalbaselineskip=11pt
\setbox\strutbox=\hbox{\vrule height8pt depth3pt width0pt}%
\let \smc=\sevenrm \let\big=\ninebig \normalbaselines
\parindent=1em
\rm}

\def\tenpoint{\def\rm{\fam0\tenrm} 
\textfont0=\tenrm \scriptfont0=\ninerm \scriptscriptfont0=\fiverm
\textfont1=\teni \scriptfont1=\seveni \scriptscriptfont1=\fivei
\textfont2=\tensy \scriptfont2=\sevensy \scriptscriptfont2=\fivesy
\textfont3=\tenex \scriptfont3=\tenex \scriptscriptfont3=\tenex
\textfont\itfam=\nineit  \def\it{\fam\itfam\nineit} 
\textfont\slfam=\ninesl \def\sl{\fam\slfam\ninesl} 
\textfont\bffam=\ninebf \scriptfont\bffam=\sevenbf
\scriptscriptfont\bffam=\fivebf \def\bf{\fam\bffam\tenbf} 
\textfont\ttfam=\tentt \def\tt{\fam\ttfam\tentt} 

\normalbaselineskip=11pt
\setbox\strutbox=\hbox{\vrule height8pt depth3pt width0pt}%
\let \smc=\sevenrm \let\big=\ninebig \normalbaselines
\parindent=1em
\rm}

\message{fin format jgr}

\hautpagegauche={\hfill\ninerm\the\auteurcourant}
\hautpagedroite={\ninerm\the\titrecourant\hfill}
\auteurcourant={R.G.\ Novikov}
\titrecourant={Formulas  for phase recovering from phaseless
scattering data at fixed frequency}

\magnification=1200
\font\Bbb=msbm10
\def\R{\hbox{\Bbb R}}
\def\C{\hbox{\Bbb C}}
\def\N{\hbox{\Bbb N}}
\def\S{\hbox{\Bbb S}}
\def\pa{\partial}
\def\b{\backslash}

\def\ep{\varepsilon}

\vskip 2 mm
\centerline{\bf Formulas for phase recovering from  phaseless  scattering data}
\centerline{\bf at fixed frequency}

\vskip 2 mm
\centerline{\bf R.G.\ Novikov}
\vskip 2 mm

\noindent
{\ninerm CNRS (UMR 7641), Centre de Math\'ematiques Appliqu\'ees, Ecole
Polytechnique,}

\noindent
{\ninerm 91128 Palaiseau, France;}

\noindent
{\ninerm IEPT RAS, 117997 Moscow, Russia}


\noindent
{\ninerm e-mail: novikov@cmap.polytechnique.fr}

\vskip 2 mm
{\bf Abstract.}
We consider quantum and acoustic wave propagation at fixed frequency for compactly supported scatterers
in dimension $d\ge 2$.
In these framework we give explicit formulas for phase recovering from appropriate  phaseless  scattering
data. As a corollary, we give  global uniqueness results for quantum and acoustic   inverse scattering
at fixed frequency without phase information.

\vskip 2 mm
{\bf 1. Introduction}

We consider the  equation
$$-\Delta\psi +v(x)\psi=E\psi,\ \ x\in\R^d,\ \ d\ge 2,\ \ E>0,\eqno(1.1)$$
where $\Delta$ is the Laplacian, $v$  is a scalar potential such that
$$\eqalign{
&v\in  L^{\infty}(\R^d),\ \ supp\,v\subset D,\cr
&D\ \ {\rm is\ an\ open\ bounded\ domain\ in}\ \ \R^d.\cr}\eqno(1.2)$$
Equation (1.1) can be considered as the quantum mechanical  Schr\"odinger equation at
fixed energy $E$.

Equation (1.1) can also be considered as the acoustic equation at
fixed frequency $\omega$. In this setting
$$E=\bigl({\omega\over c_0}\bigr)^2,\ \ v(x)=(1-n^2(x))\bigl({\omega\over c_0}\bigr)^2,\eqno(1.3)$$
where $c_0$ is a reference sound speed, $n(x)$ is a scalar index of refraction.

For equation (1.1) we consider the classical scattering solutions $\psi^+$
continuous  and bounded  on $\R^d$ and specified by the
 following asymptotics as $|x|\to\infty$:
$$\eqalign{
&\psi^+(x,k)=e^{ikx}+c(d,|k|){e^{i|k||x|}\over |x|^{(d-1)/2}}
f(k,|k|{x\over |x|})+O\bigl({1\over |x|^{(d+1)/2}}\bigr),\cr
&x\in\R^d,\ k\in\R^d,\ k^2=E,\ c(d,|k|)=-\pi i(-2\pi i)^{(d-1)/2}|k|^{(d-3)/2},\cr}\eqno(1.4)$$
where a priori unknown  function $f=f(k,l)$,
$k,l\in\R^d$,\ $k^2=l^2=E$, arising in (1.4) is the classical scattering
amplitude for (1.1).

In order to find $\psi^+$ and $f$ from $v$ one can use the following
Lippmann-Schwinger integral equation (1.5) and formula (1.7) (see, e.g., [BS], [FM]):
$$\eqalignno{
&\psi^+(x,k)=e^{ikx}+\int\limits_DG^+(x-y,k)v(y)\psi^+(y,k)dy,&(1.5)\cr
&G^+(x,k)\buildrel \rm def \over =-(2\pi)^{-d}
\int\limits_{\R^d}{e^{i\xi x}d\xi\over {\xi^2-k^2-i0}}=G_0^+(|x|,|k|),&(1.6)\cr}$$
where $x\in\R^d$,\ $k\in\R^d$,\ $k^2=E$, and $G_0^+$ depends also on $d$;
$$f(k,l)=(2\pi)^{-d}\int\limits_De^{-ily}v(y)\psi^+(y,k)dy,\eqno(1.7)$$
where $k\in\R^d$, $l\in\R^d$,  $k^2=l^2=E$.

We recall that $\psi^+$ describes scattering of the incident plane waves $e^{ikx}$ on
the potential $v$. And the second term of the right-hand side of (1.4) describes
the scattered spherical waves.

In addition to $\psi^+$, we consider also the function $R^+$ describing scattering of  spherical waves
generated by point sources. The function $R^+=R^+(x,x^{\prime},E)$, $x\in\R^d$, $x^{\prime}\in\R^d$,
can be defined as the Schwartz kernel of the standard resolvent $(-\Delta + v-E-i0)^{-1}$.
Note that $R^+=R^+(x,x^{\prime},E)=-G_0^+(|x-x^{\prime}|,\sqrt{E})$ for $v\equiv 0$, where $G_0^+$
is the function of (1.6). Given $v$, to determine $R^+$ one can use, in particular, the following
integral equation
$$R^+(x,x^{\prime},E)=-G_0^+(|x-x^{\prime}|,\sqrt{E})+\int\limits_DG_0^+(|x-y|,\sqrt{E})v(y)R^+(y,x^{\prime},E)dy,\eqno(1.8)$$
where $x\in\R^d$, $x^{\prime}\in\R^d$.

The function $R^+(x,x^{\prime},E)$ at fixed $x^{\prime}\in\R^d$ describes scattering of the spherical wave
$-G_0^+(|x-x^{\prime}|,\sqrt{E})$ generated by a point source at $x^{\prime}$. In addition,
$$\eqalign{
&R^+(x,x^{\prime},E)=-{c(d,\sqrt{E})\over (2\pi)^d}{e^{i\sqrt{E}|x|}\over |x|^{(d-1)/2}}
\psi^+\bigl(x^{\prime},-\sqrt{E}{x\over |x|}\bigr)+\cr
&O\bigl({1\over |x|^{(d+1)/2}}\bigr)\ \ {\rm as}\ \ |x|\to\infty\ \ {\rm at\ fixed}\ \ x^{\prime},\cr}\eqno(1.9)$$
$$R^+(x,x^{\prime},E)=R^+(x^{\prime},x,E),\eqno(1.10)$$
where $c$ is the constant of (1.4), $\psi^+$ is the function of (1.4), (1.5).

In connection with aforementioned facts concerning $R^+$ see, e.g., Section 1 of Chapter IV of [FM].

Let
$$\S_r^{d-1}=\{k\in\R^d:\ |k|=r\}, r>0.\eqno(1.11).$$

We consider the following three types of scattering data for equation (1.1):
\item{(a)} $f(k,l)$,  where  $(k,l)\in\Omega_f^{\prime}\subseteq\Omega_f$,
$$\Omega_f=\S_{\sqrt{E}}^{d-1}\times\S_{\sqrt{E}}^{d-1};\eqno(1.12a)$$
\item{(b)} $\psi^+(x,k)$,  where  $(x,k)\in\Omega_{\psi}^{\prime}\subseteq\Omega_{\psi}$,
$$\Omega_{\psi}=(\R^d\b (D\cup\pa D))\times\S_{\sqrt{E}}^{d-1},\eqno(1.12b)$$
assuming (1.13);
\item{(c)} $R^+(x,y,E)$,  where  $(x,y)\in\Omega_R^{\prime}\subseteq\Omega_R$,
$$\Omega_R=(\R^d\b (D\cup\pa D))\times (\R^d\b (D\cup\pa D)),\eqno(1.12c)$$
assuming (1.13).

In (1.12b), (1.12c) we assume also that
$$\R^d\b (D\cup\pa D)\ \ {\rm is\ connected}.\eqno(1.13)$$

We consider the following inverse scattering problems for equation (1.1)
 at fixed $E$:

{\bf Problem 1.1a.}
 Reconstruct potential $v$ on $\R^d$ from its scattering amplitude $f$ on some appropriate
 $\Omega_f^{\prime}\subseteq\Omega_f$.

{\bf Problem 1.1b.}
 Reconstruct potential $v$ on $\R^d$ from its scattering data
 $\psi^+$ on some appropriate
$\Omega_{\psi}^{\prime}\subseteq\Omega_{\psi}$.

{\bf Problem 1.1c.}
 Reconstruct potential $v$ on $\R^d$ from its scattering data
 $R^+$ on some appropriate
 $\Omega_R^{\prime}\subseteq\Omega_R$.

{\bf Problem 1.2a.}
 Reconstruct potential $v$ on $\R^d$ from  its phaseless scattering data
 $|f|^2$ on some appropriate
 $\Omega_f^{\prime}\subseteq\Omega_f$.

{\bf Problem 1.2b.}
 Reconstruct potential $v$ on $\R^d$ from its phaseless scattering data
 $|\psi^+|^2$ on some appropriate
$\Omega_{\psi}^{\prime}\subseteq\Omega_{\psi}$.

{\bf Problem 1.2c.}
 Reconstruct potential $v$ on $\R^d$ from its phaseless scattering data
 $|R^+|^2$ on some appropriate
 $\Omega_R^{\prime}\subseteq\Omega_R$.

Note that in quantum mechanical scattering experiments in the framework of model
described by equation (1.1) the phaseless scattering data $|f|^2$, $|\psi^+|^2$, $|R^+|^2$
of Problems 1.2a-1.2c can be measured directly, whereas the complete scattering data $f$, $\psi^+$, $R^+$
of Problems 1.1a-1.1c are not accessible for direct measurements. Therefore, Problems 1.2
are of particular interest in the framework of quantum mechanical inverse scattering.

As regards to acoustic scattering experiments in the framework of the model described by (1.1),
(1.3), the complete scattering data $f$, $\psi^+$, $R^+$ of Problems 1.2 can be measured directly.
Nevertheless, in some cases it may be more easy to measure the phaseless versions of these data.
Therefore, Problems 1.2 are also of interest in the framework of  acoustic inverse scattering.

On the other hand, in the literature many more results are given on Problems 1.1 (see [ABR],
[Be], [Bu], [BAR], [BSSR], [ChS], [E], [F1], [G], [HH], [HN], [I], [IN], [M], [Na], [N1]-[N6], [R], [S]
 and references therein) than on Problems 1.2 (see Chapter X of [ChS] and recent works [KR],
[N7] and references therein).

The works [K1], [K2], [K3], [KR], [N7] give also results on analogs of Problems 1.2, where
$E$ is not fixed. Besides, analogs of  Problems 1.2 in dimension $d=1$, where $E$ is not fixed,
were considered, in particular, in [AS], [KS].

Let
$${\cal B}_r=\{x\in\R^d:\ |x|< r\},\ \ r>0.\eqno(1.14)$$

Suppose that, for some $r>0$,
$$D\subseteq {\cal B}_r.\eqno(1.15)$$
In connection with Problems 1.1, under assumption (1.15),
it is well known, in particular, that any of the scattering data
(a) $f$ on $\Omega_f$, (b) $\psi^+$ on $\pa {\cal B}_r\times\S^{d-1}_{\sqrt{E}}$, or
(c) $R^+$ on $\pa {\cal B}_r\times \pa {\cal B}_r$ uniquely and constructively determine
two other data; see [Be].

In addition, in [N1] for $d\ge 3$ (see also [N3]) and in [Bu] for $d=2$ it was shown that $f$
on  $\Omega_f$ at fixed $E$  uniquely and constructively determines $v$ on $\R^d$, under assumption
(1.2). For related exact stability estimates, see [S], [HH], [IN], [I]. Besides, for approximate
but efficient methods for solving Problem 1.1a, see [N4], [N5], [ABR], [BAR], [N6].

The main results of the present work can be summarized as follows.

First, we give explicit asymptotic formulas for finding $f(k,l)$ at fixed $(k,l)\in\Omega_f$, $k\ne l$, from
$|\psi^+(x,k)|^2$ for $x=sl/|l|$, $s\in\Lambda= [r_1,+\infty [$ for arbitrary large $r_1\ge r$ (assuming,
e.g., (1.15)); see Theorem 2.1 and Corollary 2.1 of Section 2.

In addition, we have the asymptotic formula (2.11) for finding $|\psi^+(x^{\prime},k)|^2$ at fixed $(x^{\prime},k)\in \Omega_{\psi}$
from $|R^+(x,x^{\prime},E)|^2$ for $x=-sk/|k|$, $s\in\Lambda$ for any unbounded $\Lambda\subset [r,+\infty[$
(assuming, e.g., (1.15)).

The aforementioned  formulas give explicit reductions of Problems 1.2b, 1.2c to Problem 1.1a for appropriate
$\Omega^{\prime}_{\psi}$, $\Omega^{\prime}_R$ and $\Omega^{\prime}_f$.

In connection with reductions of Problems 1.2b, 1.2c to Problem 1.1a we give also additional global uniqueness
results summarized in Theorem 2.2 of Section 2.

Second, we give global uniqueness results for Problem 1.2b for the case when $\Omega_{\psi}^{\prime}$ is an open subset
of $\Omega_{\psi}$ and  for Problem 1.2c for the case when $\Omega_R^{\prime}$ is an open subset
of $\Omega_R$; see Theorem 2.3 of Section 2. In this connection we recall also that for Problem 1.1a in its initial formulation
 there is no uniqueness, in general; see [N7].

Actually, as soon as Problems 1.2b, 1.2c are reduced to Problem 1.1a, one can use all known results for exact
or approximate solving Problem 1.1a; see [N6] and other works cited above in connection with Problems 1.1.

Finally, we indicate some possible generalizations and extentions of results of the present work; see Remarks 2.1-2.4 at the
end of Section 2.
In particular, in a subsequent work we plan to consider phaseless inverse scattering in dimension $d=1$ when $E$
is not fixed, using an analog of Theorem 2.1 for $d=1$.

To our knowledge, no exact general result on phase recovering from phaseless scattering data for equation (1.1) at
fixed $E$ was given in the literature before the present work.

The main results of the present work are presented in detail in the next section.

\vfill\eject

\vskip 2 mm
{\bf 2. Main results}
\vskip 2 mm

We represent $f$ and $c$ of (1.4) as follows:
$$\eqalign{
&f(k,l)=|f(k,l)|e^{i\alpha(k,l)},\cr
&c(d,|k|)=|c(d,|k|)|e^{i\beta(d,|k|)}.\cr}\eqno(2.1)$$
\vskip 2 mm

We consider
$$\eqalignno{
&a(x,k)=|x|^{(d-1)/2}(|\psi^+(x,k)|^2-1),&(2.2)\cr
&a_0(x,k)=2Re\,\bigl(c(d,|k|)e^{i(|k||x|-kx)}f\bigl(k,|k|{x\over |x|}\bigr)\bigr),&(2.3)\cr
&\delta a(x,k)=a(x,k)-a_0(x,k)&(2.4)\cr}$$
for $x\in\R^d\b\{0\}$,\ \ $k\in\R^d\b\{0\}$, where $\psi^+$ are the scattering solutions of (1.4), (1.5),
$f$ is the scattering amplitude of (1.4), (1.7), $c$ is the constant of (1.4).

For real-valued potential $v$ satisfying (1.2) one can show that
$$|\delta a(x,k)|\le \delta_0(|x|,|k|),\ \ x\in\R^d,\ \ k\in\R^d\b\{0\},\eqno(2.5)$$
$$\eqalign{
&\delta_0(r,\rho)=O(r^{-1/2})\ \ {\rm for}\ \ r\to +\infty,\ d=2,\cr
&\delta_0(r,\rho)=O(r^{-1})\ \ {\rm for}\ \ r\to +\infty,\ d\ge 3,\cr}\eqno(2.6)$$
at fixed $\rho>0$ (where $\delta_0$ depends also on $v$). Estimates (2.5), (2.6) are proved in Section 3.

The key result of the present work consists in the following theorem:

\vskip 2 mm
{\bf Theorem 2.1.}
{\sl
Let real-valued potential $v$ satisfy (1.2), $d\ge 2$, and $f$, $a$ be the functions
of (1.4), (2.2). Let $(k,l)\in\Omega_f$ of (1.12a), $k\ne l$, and
$$T=2\pi\bigl(E^{1/2}\bigl(1-{kl\over E}\bigr)\bigr)^{-1}.\eqno(2.7)$$
Then the following formulas hold:
$$\eqalign{
&|f|\pmatrix{
\cos\alpha\cr
\sin\alpha\cr}=(2|c|\sin(2\pi T^{-1}(s_1-s_2)))^{-1}\times\cr
&\pmatrix{
-\sin(2\pi T^{-1}s_2+\beta)\ &\ \sin(2\pi T^{-1}s_1+\beta)\cr
-\cos(2\pi T^{-1}s_2+\beta)\ &\ \cos(2\pi T^{-1}s_1+\beta)\cr}\times\cr
&\biggl(
\pmatrix{
a((s_1+nT)l/|l|,k)\cr
a((s_2+nT)l/|l|,k)\cr}-
\pmatrix{
\delta a((s_1+nT)l/|l|,k)\cr
\delta a((s_2+nT)l/|l|,k)\cr}\biggr),n\in\N,\cr}\eqno(2.8)$$
$s_1,s_2\in [0,T]$, $s_1\ne s_2$ ($mod\, T/2$),
$\alpha=\alpha(k,l)$, $|f|=|f(k,l)|$, $|c|=|c(d,|k|)|$,
$\beta=\beta(d,|k|)$, where $\alpha$, $\beta$ are the angles of (2.1), $c$
is the constant of (1.4), $\delta a$ is defined by (2.4),
$$|\delta a(s_j+nT)l/|l|,k)|\le\delta_0(s_j+nT,|k|)=\left\{\matrix{
&O\bigl((nT)^{-1/2}\bigr),\ n\to\infty,\ d=2,\cr
&O\bigl((nT)^{-1}\bigr),\ n\to\infty,\ d\ge 3,\cr}\right.\eqno(2.9)$$
where $j=1,2$,\ $\delta_0$ is the function of (2.5).

}

Theorem 2.1 is proved in Section 3.


In connection with formula (2.8) we consider
$$\Lambda=\cup_{j=1}^{+\infty}(s_1+n_jT)\cup (s_2+n_jT),\ n_j\in\N,\ n_j\to +\infty,\ j\to +\infty,\eqno(2.10)$$
for fixed $T>0$ and $s_1,s_2\in [0,\tau]$, $s_1\ne s_2 \,(mod\,T/2)$.

\vskip 2 mm
{\bf Corollary 2.1.}
{\sl
Let real-valued potential $v$ satisfy (1.2),  $d\ge 2$, and $\psi^+$, $f$ be the scattering
functions of (1.4). Let
$(k,l)\in\Omega_f$ of (1.12a), $k\ne l$, and $T$ be defined by (2.7).
Then:


$|\psi^+(sl/|l|,k)|^2$, $s\in\Lambda$ of (2.10),  uniquely determines $f(k,l)$
via (2.1), (2.8), (2.9).

}

In addition, due to (1.9), we have that
$$|\psi^+(x^{\prime},k)|^2=(2\pi)^d|c(d,|k|)|^{-1}s^{(d-1)/2}|R^+(-sk/|k|,x^{\prime},k^2)|^2+O(s^{-1})\eqno(2.11)$$
for $s\to +\infty$ at fixed $x^{\prime}\in\R^d$, $k\in\R^d\b\{0\}$.

Formulas (2.2), (2.8),(2.9), (2.11) give explicit reductions of Problems 1.2b, 1.2c to Problem 1.1a
for appropriate $\Omega^{\prime}_{\psi}$, $\Omega^{\prime}_R$ and $\Omega^{\prime}_f$. In this connection
we have also the uniqueness results of the following theorem:

{\bf Theorem 2.2.}
{\sl
Let real-valued potential $v$ satisfy (1.2), (1.13), (1.15),  $d\ge 2$, and $\psi^+$, $f$, $R^+$ be the scattering
functions of (1.4), (1.5), (1.7), (1.8), (1.9).
Then:

(1) the scattering amplitude $f(k,l)$ for fixed $(k,l)\in\Omega_f$, $k\ne l$, is uniquely determined
by the phaseless scattering data $|\psi^+(sl/|l|,k)|^2$, $s\in [r_1,r_2]$, $r\le r_1<r_2$,

(2) the scattering amplitude $f(k,l)$ for fixed $(k,l)\in\Omega_f$, $k\ne l$, is uniquely determined
by the phaseless scattering data
$$|R^+(-sk/|k|,s^{\prime}l/|l|,E)|^2,\ \ (s,s^{\prime})\in [r_1,r_2]\times [r_1^{\prime},r_2^{\prime}],\
r\le r_1<r_2,\ r\le r_1^{\prime}<r_2^{\prime},$$
where $E$ and $r$ are the parameters of (1.12a), (1.15);

(3) the phaseless scattering data $|\psi^+|^2$ on a fixed open subset $\Omega^{\prime}_{\psi}$ of
$\Omega_{\psi}$ uniquely determine the scattering amplitude $f$ on $\Omega_f$ at fixed $E$,

(4) the phaseless scattering data $|R^+|^2$ on a fixed open subset $\Omega^{\prime}_R$ of
$\Omega_R$ uniquely determine the scattering amplitude $f$ on $\Omega_f$ at fixed $E$.

}

Theorem 2.2 is proved in Section 3. This proof involves real-analytic continuations.

As a corollary of results of [N1], [N3], [Bu] and items (3) and (4) of Theorem 2.2, we have also
the following global uniqueness results on Problems 1.2b, 1.2c:

\vskip 2 mm
{\bf Theorem 2.3.}
{\sl
Let real-valued potential $v$ satisfy (1.2), (1.13),  $d\ge 2$. Then:

$\bullet$ the phaseless scattering data $|\psi^+|^2$ on a fixed open subset $\Omega^{\prime}_{\psi}$ of
$\Omega_{\psi}$ uniquely determine $v$ in $L^{\infty}(\R^d)$,

$\bullet$ the phaseless scattering data $|R^+|^2$ on a fixed open subset $\Omega^{\prime}_R$ of
$\Omega_R$ uniquely determine  $v$ in $L^{\infty}(\R^d)$.

}

\vskip 2 mm
{\bf Remark 2.1.}
In all aforementioned results of this section the assumption that $v$ is real-valued can be replaced by
the assumption that $v$ is complex-valued and equation (1.5) is uniquely solvable for $\psi^+\in L^{\infty}(D)$
at fixed $k\in\R^d$, $k^2=E>0$.

\vskip 2 mm
{\bf Remark 2.2.}
In Problems 1.1, 1.2 the assumption that $v$ of (1.1) is compactly supported (supported in $D$)
can be replaced by the assumption that $v$ has sufficient decay at infinity; see e.g. [ChS], [E], [F1],
[G], [HN], [N1]-[N6], [R], [VW], [W], [WY].
In this case, especially in Problems 1.1b, 1.1c, 1.2b, 1.2c, it is natural to assume that $v$ is a priori
known on $\R^d\b D$. Theorem 2.1, Corollary 2.1 and formula (2.11) remain valid for $v$
with sufficient decay at infinity.

\vskip 2 mm
{\bf Remark 2.3.}
Theorem 2.1, Corollary 2.1 and formula (2.11) have analogs in dimension $d=1$.
Using these results, in a subsequent work we plan to consider phaseless inverse scattering in dimension $d=1$
when frequency is not fixed.

\vskip 2 mm
{\bf Remark 2.4.}
The approach of the present work can be also used for phaseless inverse scattering for obstacles.

\vskip 2 mm
{\bf 3. Proofs of estimates (2.5), (2.6) and Theorem 2.1}
\vskip 2 mm
{\it 3.1. Proofs of estimates (2.5), (2.6).}
Note that the asymptotic formula (1.4) holds uniformly in $x/|x|\in\S^{d-1}$. Therefore, we have that
$$\eqalignno{
&\psi^+(x,k)=\psi^+_1(x,k)+\delta \psi^+(x,k),&(3.1)\cr
&\psi^+_1(x,k)=e^{ikx}+c(d,|k|){e^{i|k||x|}\over |x|^{(d-1)/2}}f(k,|k|{x\over |x|}),&(3.2)\cr
&|\delta \psi^+(x,k)|\le\delta_1(|x|,|k|),&(3.3)\cr
&\delta_1(r,\rho)=O\bigl(r^{-(d+1)/2}\bigr)\ \ {\rm for}\ \ r\to +\infty\ \ {\rm at\ fixed}\ \ \rho>0,&(3.4)\cr}$$
where $x\in\R^d$, $k\in\R^d\b \{0\}$.

Using (2.3), (3.1), (3.2) we obtain that
$$\eqalign{
&|\psi^+(x,k)|^2=\psi^+(x,k)\overline{\psi^+(x,k)}=\cr
&1+|x|^{-(d-1)/2}a_0(x,k)+|x|^{-(d-1)}|c(d,|k|)|^2|f(k,|k|x/|x|)|^2+\cr
&2Re\,(\delta \psi^+(x,k)\overline{\psi^+_1(x,k)})+|\delta \psi^+(x,k)|^2.\cr}\eqno(3.5)$$

Due to (2.2)-(2.4), (3.5), we have that
$$\eqalign{
&\delta a(x,k)=|x|^{-(d-1)/2}|c(d,|k|)|^2|f(k,|k|x/|x|)|^2+\cr
&2|x|^{(d-1)/2}Re\,(\delta \psi^+(x,k)\overline{\psi^+_1(x,k)})+|x|^{(d-1)/2}|\delta \psi^+(x,k)|^2.\cr}\eqno(3.6)$$

Note that
$$|f(k,l)|\le C_f(\sqrt{E}),\ \ (k,l)\in\Omega_f,\eqno(3.7)$$
for some positive $C_f(\sqrt{E})$, where $\Omega_f$ is defined by (1.12a).

Using (3.3), (3.6), (3.7) we obtain (2.5) with $\delta_0(r,\rho)$ given by
$$\eqalign{
&\delta_0(r,\rho)=\delta_{0,1}(r,\rho)+\delta_{0,2}(r,\rho)+\delta_{0,3}(r,\rho),\cr
&\delta_{0,1}(r,\rho)=r^{-(d-1)/2}|c(d,\rho)|^2(C_f(\rho))^2,\cr
&\delta_{0,2}(r,\rho)=2r^{(d-1)/2}\delta_1(r,\rho)(1+c(d,\rho)r^{-(d-1)/2}C_f(\rho),\cr
&\delta_{0,3}(r,\rho)=r^{(d-1)/2}(\delta_1(r,\rho))^2,\cr}\eqno(3.8)$$
where $r>0$, $\rho>0$.

Formulas (3.4), (3.8) imply (2.6). This completes the proof of estimates (2.5), (2.6).

\vskip 2 mm
{\it 3.2. Proof of Theorem 2.1.}
Due to (2.1), (2.3), (2.7) we have that
$$\eqalign{
&a_0(sl/|l|,k)=2|c(d,|k|)||f(k,l)| \cos(2\pi T^{-1}s+\alpha(k,l)+\beta(d,|k|)),\cr
&(k,l)\in\Omega_f,\ \ k\ne l,\ \ s>0,\cr}\eqno(3.9)$$
where $\Omega_f$ is defined by (1.12a), $T$ is defined by (2.7). In addition, one can see that
$$T>0\ \ {\rm for}\ \ (k,l)\in\Omega_f,\ \ k\ne l.\eqno(3.10)$$





Due to (2.4), (3.9), we have that
$$\eqalign{
&|f|\bigl(\cos(2\pi T^{-1}s+\beta)\cos\alpha-\sin(2\pi T^{-1}s+\beta)\sin\alpha\bigr)=\cr
&(2|c|)^{-1}(a(sl/|l|,k)-\delta a(sl/|l|,k)),\ \ (k,l)\in\Omega_f,\ \ k\ne l,\ \ s>0,\cr}\eqno(3.11)$$
$\alpha=\alpha(k,l)$,\ \ $|f|=|f(k,l)|$,\ \ $|c|=|c(d,|k|)|$, $\beta=\beta(d,|k|)$.
Using (3.11) for $s=s_1+nT$ and $s=s_2+nT$, where $s_1,s_2\in [0,T]$, $n\in\N$, we obtain the system
$$\eqalign{
&\pmatrix{
\cos(2\pi T^{-1}s_1+\beta)\ &\ -\sin(2\pi T^{-1}s_1+\beta)\cr
\cos(2\pi T^{-1}s_2+\beta)\ &\ -\sin(2\pi T^{-1}s_2+\beta)\cr}
|f|\pmatrix{
\cos\alpha\cr
\sin\alpha\cr}=\cr
&(2|c|)^{-1}
\pmatrix{
a((s_1+nT)l/|l|,k)-\delta a((s_1+nT)l/|l|,k)\cr
a((s_2+nT)l/|l|,k)-\delta a((s_2+nT)l/|l|,k)\cr}.}\eqno(3.12)$$

Formula (2.8) follows from (3.12). Formula (2.9) follows from (2.5), (2.6).

Theorem 2.1 is proved.

\vskip 2 mm
{\bf 4. Proof of Theorem 2.2}

Note that
$$\eqalignno{
&|\psi^+|^2=\psi^+\overline{\psi^+},&(4.1)\cr
&|R^+|^2=R^+\overline{R^+}.&(4.2)\cr}$$

Note also that
$$\R^d\b ({\cal B}_r\cup\pa {\cal B}_r)\subseteq\R^d\b (D\cup\pa D)\eqno(4.3)$$
under assumption (1.15).

\vskip 2 mm
{\it 4.1. Proof of item (1).}
Due to equation (1.1) for $\psi^+$ and assumptions (1.2), we have that
$$-\Delta\psi^+(x,k)=E\psi(x,k),\ \ x\in\R^d\b (D\cup\pa D),\ \ {\rm for\ each}\ \ k\in\S^{d-1}_{\sqrt{E}}.\eqno(4.4)$$
Therefore,
$$\eqalign{
&\psi^+(\cdot,k)\ \ {\rm is\ (complex-valued)\ real-analytic\ on}\cr
&\R^d\b (D\cup\pa D)\ \ {\rm at\ fixed}\ \ k\in\S^{d-1}_{\sqrt{E}}.\cr}\eqno(4.5)$$

Using (4.1), (4.4) we obtain that
$$|\psi^+(\cdot,k)|^2\ \ {\rm is\  real-analytic\ on}\ \ \R^d\b (D\cup\pa D)\ \ {\rm at\ fixed}\ \ k\in\S^{d-1}_{\sqrt{E}}.\eqno(4.6)$$

As a corollary of (4.3), (4.6), we have that
$$|\psi^+(sl/|l|,k)|^2\ \ {\rm is\  real-analytic\ in}\ \ s\in ]r,+\infty [\ \ {\rm at\ fixed}\ \ (k,l)\in\Omega_f.\eqno(4.7)$$

Therefore, at fixed $(k,l)\in\Omega_f$, the function $|\psi^+(sl/|l|,k)|^2$ given for all $s\in [r_1,r_2]$,
where $r\le r_1<r_2$, uniquely determines this function for all $s\in ]r,+\infty [$ via real-analytic
continuation. This result and Corollary 2.1 imply item (1) of Theorem 2.2.

\vskip 2 mm
{\it 4.2. Proof of item (2).}
Note that
$$\eqalign{
&(-\Delta_x+v(x)-E)R^+(x,x^{\prime},E)=\delta(x-x^{\prime}),\cr
&(-\Delta_{x^{\prime}}+v(x)-E)R^+(x,x^{\prime},E)=\delta(x-x^{\prime}),\cr}\eqno(4.8)$$
where $x\in\R^d$, $x^{\prime}\in\R^d$, $E>0$, $\delta$ is the Dirac delta-function. In addition, due to (1.2), (4.8),
we have that
$$\eqalign{
&(-\Delta_x-E)R^+(x,x^{\prime},E)=\delta(x-x^{\prime}),\ x\in\R^d\b (D\cup\pa D),\ x^{\prime}\in\R^d,\cr
&(-\Delta_{x^{\prime}}-E)R^+(x,x^{\prime},E)=\delta(x-x^{\prime}),\ x\in\R^d,\ x^{\prime}\in\R^d\b (D\cup\pa D).\cr}\eqno(4.9)$$

Therefore,
$$\eqalign{
&R^+(\cdot,x^{\prime},E)\ \ {\rm  is\ (complex-valued)\ real-analytic\ on}\cr
&\R^d\b (D\cup\pa D\cup \{x^{\prime}\})\ \ {\rm for\ fixed}\ \ x^{\prime}\in\R^d,\cr
&R^+(x,\cdot,E)\ \ {\rm  is\ (complex-valued)\ real-analytic\ on}\cr
&\R^d\b (D\cup\pa D\cup \{x^{\prime}\})\ \ {\rm for\ fixed}\ \ x\in\R^d.\cr}\eqno(4.10)$$

Using (4.2), (4.10) we obtain that
$$\eqalign{
&|R^+(\cdot,x^{\prime},E)|^2\ \ {\rm  is\ real-analytic\ on}\ \ \R^d\b (D\cup\pa D\cup \{x^{\prime}\})\ \ {\rm for\ fixed}\ \ x^{\prime}\in\R^d,\cr
&|R^+(x,\cdot,E)|^2\ \ {\rm  is\ real-analytic\ on}\ \ \R^d\b (D\cup\pa D\cup \{x^{\prime}\})\ \ {\rm for\ fixed}\ \ x\in\R^d.\cr}\eqno(4.11)$$

As a corollary of (4.3), (4.11), we have, in particular, that
$$\eqalign{
&|R^+(-sk/|k|,s^{\prime}l/|l|,E)|^2\ \ {\rm is\  real-analytic\ in}\ \ s\in ]r,+\infty [\ \ {\rm if}\ \  l\ne -k\cr
&{\rm and\ in}\ \ s\in ]r,+\infty [\b s^{\prime}\ \ {\rm if}\ \ l=-k\ \ {\rm at\ fixed}\ \ s^{\prime}\in [0,+\infty [.\cr}\eqno(4.12)$$

If $l\ne -k$, then using (2.11), (4.12) one can see that the phaseless data of item (2) uniquely
determine $|\psi^+(s^{\prime}l/|l|,k)|^2$, $s^{\prime}\in [r_1^{\prime},r_2^{\prime}]$, via
real-analytic continuation  in $s$. In turn, the latter data uniquely determine $f(k,l)$ due to the
result of item (1).

Therefore, it remains to consider the case when $l=-k$. In addition, in view of (1.10), we can assume that
$r_2^{\prime}\le r_2$. Under these conditions, using (2.11), (4.12) one can see that phaseless data of
item (2) uniquely determine $|\psi^+(s^{\prime}l/|l|,k)|^2$, $s^{\prime}\in [r_1^{\prime},r_2^{\prime}[$, via
real-analytic continuation  in $s$. In turn, the latter data uniquely determine $f(k,l)$ due to the
result of item (1).

This completes the proof of item (2) of Theorem 2.2.

\vskip 2 mm
{\it 4.3. Proof of item (3).}
Note that, under our assumptions,
$$\psi^+(x,\cdot)\ \ {\rm  is\ (complex-valued)\ real-analytic\ on}\ \ \S^{d-1}_{\sqrt{E}}\ \ {\rm at\ fixed}\ \ x\in\R^d.
\eqno(4.13)$$

Actually, this result is well-known and follows from consideration of (1.5) with $k\in\C^d$, $k^2=k_1^2+\ldots +k_d^2=E$.

Using (4.1), (4.13) we obtain that
$$|\psi^+(x,\cdot)|^2\ \ {\rm  is\ real-analytic\ on}\ \ \S^{d-1}_{\sqrt{E}}\ \ {\rm at\ fixed}\ \ x\in\R^d.
\eqno(4.14)$$

Let
$$\eqalignno{
&{\cal A}_{k^{\prime},\ep}=\{k\in\S^{d-1}_{\sqrt{E}}:\ \ |k-k^{\prime}|<\ep\},\ \ k^{\prime}\in\S^{d-1}_{\sqrt{E}},\ \ \ep>0,
&(4.15)\cr
&{\cal B}_{x^{\prime},\ep}=\{x\in\R^d:\ \ |x-x^{\prime}|<\ep\},\ \ x^{\prime}\in\R^d,\ \ \ep>0.&(4.16)\cr}$$

Since $\Omega^{\prime}_{\psi}$ is open in $\Omega_{\psi}$, we can take $x^{\prime}\in\R^d\b (D\cup\pa D)$,
$k^{\prime}\in\S^{d-1}_{\sqrt{E}}$, $\ep_1>0$, $\ep_2>0$ such that
${\cal B}_{x^{\prime},\ep_1}\times {\cal A}_{k^{\prime},\ep_2}\subseteq\Omega^{\prime}_{\psi}$.
In addition, using (1.13), (4.6), (4.14) one can see that already $|\psi^+|^2$ on
${\cal B}_{x^{\prime},\ep_1}\times {\cal A}_{k^{\prime},\ep_2}$ uniquely determines $|\psi^+|^2$ on $\Omega_{\psi}$
via sequential real-analytic continuations in $x$ and in $k$.

In turn, $|\psi^+|^2$ on $\Omega_{\psi}$ uniquely determines $f$ for all $(k,l)\in\Omega_f$, $k\ne l$, due to
Corollary 2.1. In addition, under our assumptions, $f$ is continuous on $\Omega_f$ and therefore, is
uniquely determined also for $k=l$.

This completes the proof of item (3) of Theorem 2.2.

\vskip 2 mm
{\it 4.4. Proof of item (4).}
Since $\Omega^{\prime}_R$ is open in $\Omega_R$, we can take $y\in\R^d\b (D\cup\pa D)$,
$y^{\prime}\in\R^d\b (D\cup\pa D)$, $\ep>0$, $\ep^{\prime}>0$ such that
${\cal B}_{y,\ep}\times {\cal B}_{y^{\prime},\ep^{\prime}}\subseteq\Omega^{\prime}_R$.
In addition, using (1.13), (4.11) one can see that already $|R^+|^2$ on ${\cal B}_{y,\ep}\times {\cal B}_{y^{\prime},\ep^{\prime}}$
uniquely determines $|R^+(x,x^{\prime},E)|^2$ for all $(x,x^{\prime})\in\Omega_R$, $x\ne x^{\prime}$,
via sequential real-analytic continuations in $x$ and in $x^{\prime}$.
In turn, the latter data uniquely determine $|\psi^+|^2$ on $\Omega_{\psi}$
 via (2.11). Finally, due to item (3), $|\psi^+|^2$ on $\Omega_{\psi}$  uniquely determines $f$ on $\Omega_f$.

This completes the proof of item (4) of Theorem 2.2.

\vskip 4 mm
{\bf Acknowledgment.}
The author is very grateful to M.V. Klibanov for remarks that have helped to simplify considerably  the formulations
of Theorem 2.1 and Corollary 2.1 of the first version (see [N8]) of the present article.

\vfill\eject

\vskip 4 mm
{\bf References}
\vskip 2 mm
\item{[ AS]} T. Aktosun, P.E. Sacks, Inverse problem on the line without phase information,
Inverse Problems 14, 1998, 211-224.
\item{[ABR]} N.V. Alexeenko, V.A. Burov, O.D. Rumyantseva, Solution of the
 three-dimensional
acoustical inverse scattering problem. The modified Novikov
algorithm, Acoust. J. 54(3), 2008, 469-482 (in Russian), English transl.:
Acoust. Phys. 54(3), 2008, 407-419.
\item{[ Be]} Yu.M. Berezanskii, On the uniqueness theorem in the inverse problem of
spectral analysis for the Schr\"odinger equation, Tr. Mosk. Mat. Obshch. 7, 1958, 3-62 (in Russian).
\item{[ BS]} F.A. Berezin, M.A. Shubin, The Schr\"odinger Equation,
Vol. 66 of Mathematics and Its Applications, Kluwer Academic, Dordrecht, 1991.
\item{[ Bu]} A.L. Buckhgeim, Recovering a potential from Cauchy data in the
two-dimensional case, J. Inverse Ill-Posed Probl. 16(1), 2008,  19-33.
\item{[BAR]} V.A. Burov, N.V. Alekseenko, O.D. Rumyantseva, Multifrequency
generalization
of the Novikov algorithm for the two-dimensional inverse scattering
problem, Acoust. J. 55(6), 2009, 784-798 (in Russian); English transl.:
 Acoustical Physics 55(6), 2009,  843-856.
\item{[ChS]} K. Chadan, P.C. Sabatier, Inverse Problems in Quantum Scattering
Theory, 2nd edn. Springer, Berlin, 1989
\item{[  E]} G. Eskin, Lectures on Linear Partial Differential Equations,
Graduate Studies in Mathematics, Vol.123, American Mathematical Society, 2011.
\item{[ F1]} L.D. Faddeev, Uniqueness of the solution of the inverse
scattering problem, Vest. Leningrad Univ. 7, 1956, 126-130 [in Russian].
\item{[ FM]} L.D. Faddeev, S.P. Merkuriev, Quantum Scattering Theory for Multi-particle
Systems, Nauka, Moscow, 1985 (in Russian); English transl: Math. Phys. Appl. Math. 11 (1993),
Kluwer Academic Publishers Group, Dordrecht.
\item{[  G]} P.G. Grinevich, The scattering transform for the two-dimensional
Schr\"odinger operator with a potential that decreases at infinity at fixed
nonzero energy, Uspekhi Mat. Nauk 55:6(336),2000, 3-70 (Russian); English
translation: Russian Math. Surveys 55:6, 2000, 1015-1083.
\item{[ HH]} P. H\"ahner, T. Hohage, New stability estimates for the inverse
acoustic inhomogeneous
medium problem and applications, SIAM J. Math. Anal., 33(3),
2001, 670-685.
\item{[ HN]} G.M. Henkin, R.G. Novikov, The $\bar\partial$-equation in the
multidimensional
inverse scattering problem, Uspekhi Mat. Nauk 42(3), 1987, 93-152
(in Russian);
English transl.: Russ. Math. Surv. 42(3), 1987, 109-180.
\item{[ I]} M.I. Isaev, Energy and regularity dependent stability estimates for near-field
inverse scattering in multidimensions, Journal of Mathematics, Hindawi Publishing Corp.,
2013, DOI:10.1155/2013/318154.
\item{[IN]} M.I. Isaev, R.G. Novikov, New global stability estimates for
monochromatic
inverse acoustic scattering, SIAM J. Math. Anal. 45(3), 2013, 1495-1504
\item{[ K1]} M.V. Klibanov, Phaseless inverse scattering problems in three dimensions,
SIAM J. Appl. Math. 74, 2014, 392-410.
\item{[ K2]} M.V. Klibanov, On the first solution of a long standing problem:
uniqueness of the phaseless quantum inverse scattering problem in 3-d,
Appl. Math. Lett. 37, 2014, 82-85.
\item{[ K3]} M.V. Klibanov, Uniqueness of two phaseless non-overdetermined inverse acoustic
problems in 3-d, Applicable Analysis 93, 2014, 1135-1149.
\item{[ KR]} M.V. Klibanov, V.G. Romanov, Reconstruction formula for a 3-d phaseless inverse
scattering problem for the Schr\"odinger equation, arXiv:1412.8210v1, December 28, 2014.
\item{[ KS]} M.V. Klibanov, P.E. Sacks, Phaseless inverse scattering and the phase problem in
optics, J.Math. Physics 33, 1992, 3813-3821.
\item{[  M]} R.B. Melrose, Geometric scattering theory, Cambridge University Press, 1995.
\item{[ Na]} A.I. Nachman, Reconstructions from boundary measurements, Ann. Math. 128, 1988, 531-576.
\item{[ N1]} R.G. Novikov, Multidimensional inverse spectral problem for the
equation
$-\Delta\psi+(v(x)-Eu(x))\psi=0$, Funkt. Anal. Prilozhen. 22(4), 1988, 11-22
(in Russian); English transl.: Funct. Anal. Appl. 22, 1988, 263-272.
\item{[ N2]} R.G. Novikov, The inverse scattering problem at fixed energy
level for the two-dimensional Schr\"odinger operator, J. Funct. Anal., 103,
1992, 409-463.
\item{[ N3]} R.G. Novikov, The inverse scattering problem at fixed energy for
Schr\"odinger equation with an exponentially decreasing potential,
Comm. Math.
\item{     } Phys. 161, 1994,  569-595.
\item{[ N4]} R.G. Novikov, Approximate inverse quantum scattering at fixed energy in dimension 2,
Proc. Steklov Inst. Math. 225, 1999, 285-302.
\item{[ N5]} R.G. Novikov, The $\bar\partial$-approach to monochromatic inverse
 scattering in three dimensions, J. Geom. Anal. 18, 2008, 612-631.
\item{[ N6]} R.G. Novikov, An iterative approach to non-overdetermined inverse scattering at
fixed energy, Mat. Sb. 206(1), 2015, 131-146 (in Russian).
\item{[ N7]} R.G. Novikov, Explicit formulas and global uniqueness for phaseless inverse
scattering in multidimensions, J. Geom. Anal. DOI:10.1007/s12220-014-9553-7;
\item{     }  arXiv:1412.5006v1, December 16, 2014.
\item{[ N8]} R.G. Novikov, arXiv:1502.02282v1, February 8, 2015.
\item{[  R]} T. Regge, Introduction to complex orbital moments, Nuovo Cimento 14, 1959,
951-976
\item{[  S]} P. Stefanov, Stability of the inverse problem in potential
scattering at fixed
energy, Annales de l'Institut Fourier, tome 40(4), 1990, 867-884.
\item{[ VW]} A.Vasy, X.-P. Wang, Inverse scattering with fixed energy for
dilation-analytic potentials, Inverse Problems 20, 2004, 1349-1354.
\item{[  W]} R. Weder, Global uniqueness at fixed energy in multidimensional inverse scattering
theory, Inverse Problems 7, 1991, 927-938.
\item{[ WY]} R. Weder, D. Yafaev, On inverse scattering at a fixed energy
for potentials with a regular behaviour at infinity,
Inverse Problems 21, 2005, 1937-1952.

\end